\documentclass[]{article}
\usepackage{amsmath,amssymb,bm,bbm,mathrsfs,amscd}
\usepackage{calc}
\usepackage{color}
\usepackage{graphicx}
\usepackage{epstopdf}
\usepackage{epsfig}
\usepackage{tikz}
\usepackage{hyperref}
\usepackage{amssymb}
\usepackage{physics}
\usepackage{braket}
\newtheorem{definition}{Definition}
\newtheorem{proposition}{Proposition}
\title{Exclusion of Excluded Middle Law and Quantum Mechanics}
\author{Maziar Esfahanian and Lodewijk Arntzen}
\begin{document}
\maketitle

\abstract{\noindent In this paper we discuss limitions appearing while using Boolean Logic in Non-Constructive Mathematics (NCM) as a mathematical foundation for Quantum Mechanics. These limitations vanish naturally using Constructive Mathematics (CM) and Topos Theory which is one of its models. The line of reasoning is in agreement with recent theoretical work on Quantum Logic. In our discussion we address the pivotal role of the excluded middle law in more detail.}

\maketitle
\section*{Introduction}\label{sec1}
Quantum Mechanics has deeply changed the way we view our world. Already in the early days of Quantum Mechanics, debates between Einstein and Bohr \cite{bohr} revealed remarkable consequences following from this theory, with one of the central questions: is  Quantum Mechanics compatible with Local Realism? The exploration of all consequences of Quantum Mechanics already took a century, and this process continues up until the present day. In this paper we will obviously not discuss the complete history of Quantum Mechanics and all available interpretations. We decided to start with some illustrative milestones. Schr\"{o}dinger coined the term 'Verschr\"{a}nkung' (Entanglement) directly after the famous 1935 paper from Einstein, Podolsky and Rosen appeared \cite{EPR}. The intention of this paper was to show that Quantum Mechanics could not be complete. John Bell continued in 1964, stating two basic common sense assumptions concerning our physical world, 1. only local interactions are allowed, i.e. no interactions can exceed the speed of light (locality). 2. All objects contain physical properties independent of the observer \cite{Bell} (realism). These assumptions were fully reasonable at the time, in fact defining local realism, and leading directly to the famous Bell inequalities. At present (2023) these inequalities (although the original inequalities were adapted for better experimental access) are now known to be violated beyond any reasonable doubt using delicate experiments with entangled photons\footnote{Nobel prize 2022, Aspect, Clauser and Zeilinger}. Four decades earlier, in 1983, Feynman made a huge step forward in understanding the full impact of Quantum Mechanics \cite{Feynman}. In a famous thought experiment, he brilliantly played with the idea of simulating a confined part of space and time using a classical computer only, and the computing power scales with the confined volume. Assuming nature to obey classical physics laws, an infinitely correct simulation is concluded to be impossible, this would for instance lead to classical EM waves with infinite wavelengths. Realizing that the world is not classical at all does not make things easier. Due to the violation of Bell's inequalities, basic local realism assumptions are untenable, and a infinitely correct simulation using any classical computer of a confined piece of space-time is now completely out of reach. Conclusion: a quantum system can only be simulated by another quantum system. Any universal classical computer fails to fully represent a quantum system. In fact, while recognizing this already in the 1980s, Feynman directly coined the idea of a quantum computer. Important for our discussion here is that, quantum phenomena, such as superposition and entanglement, seem to be fundamentally different and are principally and logically impossible to describe using classical logic. To illustrate this with one concrete example: a superconducting current running clockwise and counterclockwise at the same time is classically out of the question since the classical line of reasoning is automatically Boolean, and in this manner we are logically forcing ourselves to make a clear choice: the current is clockwise \textit{or} the current is counterclockwise. zero \textit{or} one. Following the same classical reasoning, a qubit can logically not exist. The idea, that the logic behind quantum mechanical systems must be different in principle, was already recognized in 1935, only two years after the mathematical formalism of quantum mechanics had been fully developed. In 1933 Von Neumann published his famous “Mathematische Grundlagen der Quantenmechanik” \cite{vonneumann1}. This book still serves as a mathematical foundation for quantum mechanics up until the present day. Despite the wide acceptance of this book, Von Neumann and Birkhoff already posed several intriguing questions two years later. In their paper “The Logic of Quantum Mechanics \cite{vonneumann2}” Birkhoff and Von Neumann express the idea that 'The Principle of Non-Commutativity of Observations' directly reveals that the logic behind Quantum Mechanics cannot be Boolean. From a Set Theoretical point of view, the underlying logic of the calculus needed to support Quantum Mechanics requires a richer type of logic. Although this notion of von Neumann and Birkhoff is indeed quite a warning, already expressed in 1935, it did not receive much attention at the time. Andreas D\"{o}ring and Chris Isham \cite{doring} constructed a topos theoretical framework for quantum mechanics. Even more recently, the development of topos theory in relation with quantum logic has been developed considerably and has at present reached an established status with the appearance of the book 'Foundations of Quantum Theory' of Klaas Landsman \cite{Landsman2017} in 2017. In  this paper, we explore some consequences of this notion and will explore the history of calculus first, resulting into two branches, Non-Constructive Mathematics (NCM) and constructive mathematics (CM). We agree to the position that Constructive Mathematics and Topos Theory, which is one of its models, fits more naturally to the logic behind Quantum Mechanics. In short: a calculus based on Boolean logic is only sufficient for classical physics and classical computing - and we will explain and construct our position leading to a final conclusion. 
\subsection*{Quantum Mechanics, CM and NCM}
 So why is Non-Constructive Mathematics (NCM) more suitable as a mathematical foundation for Quantum Mechanics? In order to answer this question, we discuss several notions of Constructive Mathematics (CM) and distinguish them from Non-Constructive Mathematics (NCM). An historical overview in not feasible within the scope of this paper. Therefore, we decided to describe the very heart of CM only. The description is considered sufficient to make the line of reasoning understandable. The next step is to explain the reason why CM is needed to develop a consistent mathematical foundation of Quantum Mechanics and explain why non-constructive mathematics (NCM) is not suitable. We point out here for clarity (and maybe also for reassurance) that all well known and well established mathematical tools (i.e. Hilbert Space, linear Hermitian operators, inner product) find support without any problem within CM. 
 \section*{Calculus Approaches}
In order to discuss how Boolean logic enters calculus, we need to go back to the 17$^{\mathrm{th}}$ century. It is widely known that Newton and Leibniz influenced mathematics in that century deeply. Without these developments, almost none of our achievements in astronomy, aerospace science, mechanical engineering, modern medicine and many other fields of science would have been possible. Surprisingly, far less widely known is the fact that the two calculus approaches differ considerably. A noticeable difference is found in the notion of the \emph{infinitesimal}. In the Newtonian approach, we have the common notion of \emph{limit} and the elements of calculus are built based upon this notion. This limit concept is constructed in a somewhat unnatural way\footnote{In reference Differential Geometry in Toposes\cite{Kostecki} this is described in more detail.}. The Leibniz approach is considered more natural, both logically and synthetically. We will now explain why this approach may hold the key to a more suitable mathematical foundation for Quantum Mechanics. In order to explain this, we first need to introduce an elementary, but formal definition of the notion of infinitesimal, just sufficient for our discussion within the scope of this paper. 
\begin{definition}
Suppose that $R$ is a commutative ring. We set:
$$D:=\{ x \in R \vert x^{2}=0 \}.$$ The elements of the set $D$ are called \textbf{infinitesimals}.
\end{definition}
\noindent More precisely, these elements are called \emph{first order infinitesimals}. From an algebraic point of view, the elements of $D$ are \emph{nilpotent} elements of the ring $R$. It is clear that $D\subset R$. In the Leibniz approach we assume that $D\neq\{0\}$, thus $R$ can not be regarded as the field of real number $\mathbb{R}$. Roughly speaking, $D\neq \{0\}$ guarantees the \emph{existence} of some elements in $R$ like $x$ which are not equal to zero, but are extremely small such that: $x^{2}=0$. There are several philosophical arguments about the existence of infinitesimals among mathematicians, even contemporary mathematicians. Among them, the ideas of French Fields medalist Alain Connes are important, due to the connection of these ideas with Non Commutative Geometry (NCG). We refer the reader to \cite{Conneshyper} for detailed discussions about this recent issue. Differential Geometry (DG) is based on the notion of \emph{smoothness}, and more primarily on the notion of \emph{limit} in calculus and represents a well known branch of mathematics. From this point of view it belongs to the Newtonian approach. However, from another point of view, which is related to the notion of space with \emph{locally Euclidean} features instead of \emph{globally Euclidean}, DG does not belong to the Newtonian school. DG is the most important geometrical tool to study general relativity (GR). Furthermore, DG beside other mathematical tools, like category theory, cobordisms theory, algebraic geometry and homological algebra, create the mathematical framework of research on both approaches to quantum gravity, which are string theory and loop quantum gravity. In the following, using the notion of infinitesimal (instead of using the notion of limit), we introduce the main axiom of synthetic differential geometry (SDG) which is the analogue of DG in the Leibniz approach. This axiom is called the Kock-Lawvere axiom, and is formulated as follows: 
\begin{equation}
\forall g \in R^{D}, \exists !b\in R, \forall d \in D : \quad g(d)=g(0)+d \cdot b,
\end{equation}
where $ g \in R^{D}$ means $g$ is a function from $D$ to the $R$. In words, statement 1 means: for each function $g$ from $D$ to the $R$, there exists a unique element of $R$ like $b$, such that for each $d$ belonging to $D$, we have $g(d) = g(0)+ d \cdot b$. More details on this can be found in the reference Differential Geometry in Toposes \cite{Kostecki}. At this stage in our discussion, we need to introduce the \emph{law of excluded middle}, to see its pivotal role in shaping the two different mathematical approaches radically. We argue that it is necessary to exclude this law, if we want the calculus approach to serve as a foundation for Quantum Mechanics.
\begin{definition}\label{LEM}
\textbf{The Law of Excluded Middle} (LEM): For any logical proposition $p$
either $p$ is true or $\neg p$ is true. By $\neg p$ we mean the negation of $p$. 
\end{definition}
It is immediately seen from this that LEM stems from a \emph{Boolean logic} point of view, spectacles of \emph{zero or one, to be or not to be, alive or dead, clockwise or counterclockwise}. Exactly this law forces the logic behind the calculus approach to be Boolean. We shall discuss the consequences of this later in more detail.
\begin{proposition}
The Kock-Lawvere axiom is not compatible with the law of
excluded middle.
\end{proposition}
For a proof see \cite{Kostecki}. This theorem suggests that we should exclude LEM from our mathematical foundation if we want to be permitted to talk about infinitesimals. We need $D$ not to contain only zero. We need infinitesimals in order to use the \emph{Kock-Lawvere} axiom. Hence, in the Leibniz approach, the law of excluded middle is not valid. 
\begin{proposition}
The axiom of choice implies the law of excluded middle (LEM). 
\end{proposition}
We refer the reader to \cite{Kostecki} for a proof. Proposition 2 suggests yet another exclusion. In the framework of constructive mathematics and the Leibniz approach we reject the validity of the axiom of choice.

\subsection*{The Advantage of the Leibniz Approach}
Let us address the central question: Why does the Leibniz approach lead to a more suitable mathematical foundation for Quantum Mechanics? And why is the Newtonian approach problematic? Here, the Law of Excluded Middle (definition \ref{LEM}) plays a pivotal role. Let us first recall the axioms of Quantum Mechanics:
\begin{itemize}
\item[Axiom 1.] We assume that a Hilbert space is associated with any isolated physical system and we call this Hilbert space, \emph{state space of the system}. A system is completely described by its state vector, which is a unit vector in the state space of the system. 
\item[Axiom 2.] The evolution of an isolated (closed) quantum system is described by a unitary transformation. The time evolution of the state of an isolated quantum system is described by the Schr\"{o}dinger's equation as follows: 
\begin{equation}
i \hslash \frac{d \ket{\psi}}{dt} =H \ket{\psi}.
\end{equation}
In this equation $\hslash$ is a physical constant known as Planck’s constant. 
\item[Axiom 3.] Quantum measurements are described by a collection $M_{m} $ of measurement operators. The index $m$ refers to the measurement outcomes that may occur in the experiment. These are operators acting on the state space of the system being measured. If the state of the quantum system is $ \ket{\psi}$, immediately before the measurement then the probability that result $m$ occurs is given by
\begin{equation}
p(m)=\bra{\psi}M_{m}^{\dagger} M_{m}\ket{\psi}
\end{equation}
where $ p(m)=\bra{\psi}M_{m}^{\dagger} M_{m}\ket{\psi} $ is inner product of $\ket{\psi}$ and $M_{m}^{\dagger} M_{m}\ket{\psi} .$
Since according to the axioms of probability theory, a quantum system must be in one of the possible states after the act of measurement (for instance, a dice after dropping should be in one state between 1 and 6) the measurement operators should satisfy the completeness condition: 
\begin{equation}
\sum_{m} p(m)=\sum_{m}\bra{\psi}M_{m}^{\dagger} M_{m}\ket{\psi}=1 .
\end{equation}
Axiom 3 is at the root of various interpretations of quantum mechanics, since it shows the role of the rest of world (with respect to the closed quantum system) i.e. the environment, and its interaction with the quantum system which includes the measurement process performed by the observer.
\item[Axiom 4.]\emph{The superposition principle}. This principle states that if $\ket{x}$ and $\ket{y}$ are two states of a quantum system, then any \emph{superposition} $\alpha\ket{x}+\beta\ket{y}$ should also be an allowed state of the quantum system, where $\alpha^2 + \beta^2 =1$. See \cite{Nielsen,Benenti}.
\end{itemize}
These axioms represent the common starting point of interesting and intensive debates on the interpretation of quantum mechanics. We assume that the reader is familiar with the most well known interpretation schools\footnote{Among the well known schools: Copenhagen Interpretation (CI), Many World Interpretation (MWI), De Broglie-Bohm interpretation and Quantum Bayesianism. It may be unnecessary to point out that all schools are refering to the same axioms and the same mathematical framework.}. Within the scope of this paper, of course, we cannot go into detail with respect to these interpretations, but we will pick out issues that are of importance for our discussion here. One important debate starts naturally while interpreting Axiom 2, implying that the time evolution of closed system is in fact completely deterministic. Interpreting this axiom in isolation of other axioms makes it quite hard to see why quantum mechanics is connected to in-determinism in such a fundamental way. The possible in-determinism in interpretation stems from Axiom 3, leading in the Copenhagen interpretation to the well known \emph{measurement problem} and the \emph{collapse of the wave function}. Since the violation of Bell's inequalities, the validity of local realism is seriously doubted. This development supports the idea that quantum mechanics differs logically from classical physics and motivates us to explore this idea in more detail. In classical physics, no fundamental reason is found preventing a prediction of the outcome of a future measurement of any quantity of a closed system. In our view, in agreement with a common interpretation, Axiom 3 shows that fundamentally no fixed computable prediction of the outcome of measurements is possible. In fact, axiom 3 introduces a new kind of unpredictability that differs fundamentally from the well know unpredictability found in classical theories like classical Chaos Theory or Thermodynamics and Statistical Physics. This classical unpredictability in its logic remains completely deterministic in nature, and it refers to the extreme sensitivity of the system to its initial conditions, or it refers to the lack of knowledge regarding the state of the system. In quantum mechanics however, in this common interpretation, the measurement 'act', fundamentally prevents a deterministic prediction of the outcome of the experiment. Let us now focus on the $4^{\mathrm{th}}$ axiom, the \emph{superposition principle}. This axiom seems counter-intuitive at a first glance, and here indeed the problem with the Newtonian approach starts. In the Newtonian approach, the law of excluded middle is held, hence in Schr\"{o}dinger's cat scenario, we pick up the following oppositional states $\ket{\mathrm{alive}}$ and $\ket{\mathrm{dead}}$, implying there is no valid state in a Newtonian context. In other words, the underlying logic of the Newtonian school of mathematics, and more specifically, the law of excluded middle, is inconsistent with the logic needed for the superposition concept. Indeed, the state of the cat should indeed really be allowed to be dead and alive at the same time. Of course, it is perfectly possible to write a simple linear combination of two states within the Newtonian approach. But this approach cannot support the logic needed for the full support of the notion of superposition. The underlying logic of the Newtonian approach is in its core Boolean, and the logic behind superposition really seems to require a non-Boolean approach. In contrast, the Leibniz approach of mathematics is not restricted by this logic, and remains fully consistent to the notion of superposition. In this approach the law of excluded middle is not valid, enabling us to write down a superposed state $\ket{\mathrm{alive}}+\ket{\mathrm{dead}}$ for Schr\"{o}dinger's cat in its full meaning. This superposition is crucial for quantum behavior, it allows superconducting currents to run clockwise and counterclockwise at the same time, it allows a spin $\frac{1}{2}$ particle to be up and down at the same time, and in all possible positions in between, and allows even a qubit to exist. Boolean logic forces the system to be up \textit{or} down, and excludes all positions in between, in fact directly destroying quantumness. In classical physics, this is simply is not a problem, since a superposition does require any positions in between. Referring back to the Feynman thought experiment: we argue that it is this richer logic that distinguishes a quantum computer from a classical computer. This seems to be one of the basic reasons why a classical computer principally fails to represent a quantum system. 
From a \emph{"toposopher's"}\footnote{The term \emph{"toposopher"} for the first time is suggested by the famous algebraic geometer \emph{Miles Ried}. See \cite{Johnstone} } point of view, this fact is related to the internal logic of a chosen \emph{topos} which is not equivalent with the topos of sets. In other words, the Newtonian approach is based on the Boolean topos of sets, while the Leibniz approach is based on a considerably richer non-Boolean topos. We refer the interested reader to\cite{Johnstone} to learn more about topos theory. 
\section*{Conclusion}
In this paper we argue and support the idea that Quantum Mechanics and its associated phenomena, such as superposition and entanglement, require a underlying logic richer than the standard Boolean logic. If we explore the supporting calculus and its history, we find that the Newton approach indeed is embedded in Boolean logic, and here, the Law of Excluded Middle (LEM) plays a pivoting role. The Leibniz approach to calculus shows that this law may not be essential, a valid mathematical framework is still possible. Without this law of excluded middle (LEM), it is still perfectly possible to represent all the usual mathematical tools needed for Quantum Mechanics, including Hilbert Space and all the tools associated with it, and numerically the calculations will lead to exactly the same answer. 
We also point out that we do not wish to change any of the (von Neumann) axioms of Quantum Mechanics. However, this all does not mean that the underlying logic may be ignored, i.e. one can try to ignore this and still continue, since numerically there will be no difference. Problematic situations and paradoxes will appear due to this non fitting logic, in fact, they have already appeared. One class of them concerns problems in the foundations of string theory. T-duality forced some physicists to think about more counter-intuitive ideas such as mirror symmetry. We recall that one of the most well-known and applicable models of constructive mathematics is topos theory. Developing a topos theoretical framework for Quantum Mechanics has been the topic of work from Andreas D\"{o}ring and Chris Isham \cite{doring}, and more recently, the development of Topos Theory and Quantum Logic has entered a new established status after the appearance of the book 'Foundations of Quantum Theory' (Klaas Landsman) in 2017\cite{Landsman2017} . Exploring the mathematics in itself that is used to describe physics has led the way to new physics many times.  One illustrative example from the past: from classical electrodynamics, the displacement current was predicted to exist, since otherwise the mathematics would not be consistent. Analogously, we expect to deepen understanding of quantum mechanics further by exploring the logic behind Quantum Mechanics including the mathematics behind it.
\subsection*{Conflict of Interest Statement}
On behalf of all authors, the corresponding author states that there is no conflict of interest.

\end{document}